\documentclass[10pt,twoside]{epnt1p}
\usepackage{epsfig}

\usepackage{amssymb}
\usepackage{amsmath}

\usepackage{url}

\begin{document}

\begin{frontmatter}


\title{Direct Detection of Physics Beyond the Standard Model}


\author[address1]{Ivone F. M. Albuquerque}

\address[address1]{Instituto de F\'{i}sica, Universidade de S\~{a}o Paulo, Brasil}

\begin{abstract}
In supersymmetric theories where the lightest supersymmetric particle is the
gravitino the next to lightest supersymmetric particle (NLSP) is typically a 
long lived charged slepton. These NLSPs can be produced by
high energy neutrinos interactions with nucleons in the Earth and be detected
by km$^3$ neutrino telescopes. The signal, consists of two parallel charged 
tracks separated by a few hundred
meters. This is compared to the main background, coming from
direct di-muon production. The distance between the background tracks is much
smaller and allows for a clean separation from the NLSP signal.
We conclude that neutrino telescopes
will complement collider searches in the determination of the supersymmetry 
breaking
scale, and may even provide the first evidence for supersymmetry at the weak scale.
\end{abstract}


\end{frontmatter}

\section{\label{sec:intro} Introduction}
One of the most attractive candidate theory for the extension of the
standard model of particle physics is weak scale supersymmetry (susy).
The particle spectrum of this model is determined by the susy breaking
mechanism ($\sqrt{F}$). When susy is broken at high scales
such that $\sqrt{F} \gtrsim 10^{10}$GeV the LSP is typically the 
neutralino. If however susy is broken at lower scales, 
$\sqrt{F} \lesssim 10^{10}$GeV, the LSP
tends to be the gravitino and the Next to
Lightest Supersymmetric Particle (NLSP) is usually a charged slepton, typically the
right-handed stau. Within these models, if $\sqrt{F}$ is much
larger than a TeV the stau NLSP decay is suppressed \cite{abcl}
and its lifetime can be very large.

It was recently proposed \cite{abcl} that the diffuse flux of high
energy neutrinos colliding with the Earth can produce pairs of slepton NLSPs.
The energy loss of these particles is very small and they travel for long distances
before stopping. This compensates their small production cross section 
since they can be detected far away from their production point.

The NLSP production typically has a 
high boost and they will not decay inside the
Earth provided the susy breaking scale is $\sqrt{F} > 10^7$GeV.
Since the NLSP is charged, its upward going tracks can be
detected in large ice or water Cerenkov detectors, such as IceCube.  

We performed a Monte Carlo simulation of the NLSP production and
propagation through the Earth and determined their signature in km$^3$ 
neutrino telescope 
\cite{abc2}. The main  background (di-muon events) was also simulated and 
compared to the NLSP signal.

\section{\label{sec:prod} NLSP Production and Propagation Through the Earth}

The susy processes for NLSP production is analogous to the
standard model (SM) charged current (CC) interaction and involves a t-channel
production of a left-handed slepton ($\tilde l_L$) and a squark ($\tilde q$) 
through a gaugino exchange. In the dominant process the gaugino is a chargino 
and in the subdominant process a neutralino.
The $\tilde l_L$ and $\tilde q$ prompt decay results in two lighter
right-handed sleptons (NLSPs) plus SM particles. 
The NLSP will always be produced in pairs and is typically 
the stau ($\tilde\tau_R$). 

The energy threshold
for the NLSP production is given by the $\tilde l_L$ and $\tilde q$
masses. We take the typical values of 250 GeV for both chargino and
$\tilde l_L$ masses, 150 GeV for the NLSP mass and set the $\tilde q$  mass to
three values of 300, 600 and 900 GeV.

We compute the NLSP cross section including both dominant and subdominant
processes. The susy cross section is about three orders of magnitude
lower than the SM production cross section \cite{abc2}. 

For the high energy neutrino flux reaching the Earth, we take the 
Waxman-Bahcall~\cite{wb} limit

\begin{equation}
\left(\frac{d\phi_\nu}{dE}\right)_{\rm WB} = \frac{(1-4) \times
10^{-8}}{E^2} {\rm GeV~ cm^{-2} s^{-1} ~sr{-1}}~.
\label{eq:wblimit}
\end{equation}
Since the NLSP production
is independent of the neutrino flavor, the initial
neutrino flux contains both $\nu_\mu$ and $\nu_e$ in a 2 : 1 ratio.

The NLSP propagation through the Earth depends on its energy loss which at 
high energies is dominated by radiative losses.
The main processes are bremsstrahlung, pair production and photo-nuclear
interactions~\cite{abc2,ina}. As a result the NLSP will travel distances
much greater than a muon which will compensate for the lower cross section.

\section{Simulation of NLSP Production, Propagation and Detection}

In order to analyze the NLSP production, propagation and detection
we developed a Monte Carlo simulation. Assuming an isotropic incoming
neutrino flux we generated 30K NLSP events for each $\tilde q$ mass.
The neutrino incoming energy is distributed in steps ranging from the stau
production threshold to $10^{11}$ GeV. 

The survival probability ($P_S$) for a neutrino traveling through a 
path $dl$ is given by $P_S = \exp(\int n \sigma dl)$
where n is the Earth number density \cite{gqrs} and $\sigma$ is the
interaction cross section. The probability of interaction is $1 - P_S$.

To simulate the NLSP production an interaction point is randomly chosen
from an interaction probability distribution. If the interaction point falls
within a distance from the detector that is smaller than the NLSP range
the event is accepted. If this distance is greater than the NLSP range
the event will only count for the normalization.

The center of mass (CM) angular distribution was chosen based on the 
differential cross section distribution. We assume that the CM angular
distribution of the right-handed slepton pair (NLSPs) is the same as
the CM angular distribution of the $\tilde l_L$ and $\tilde q$ which is
a good approximation~\cite{abc2}. The 4-momentum of the NLSP pair is determined 
in the CM from the angular distribution and boosted to
the lab frame. The lab angular ($\Theta_{lab}$) distribution is determined 
from the lab 4-momentum.

With this procedure we can determine not only the rate of events at the
detector as well as the two NLSP track separation in the detector.
The separation is simply $\Theta_{lab}$ times the distance
from the neutrino interaction point and the detector.

\section{Background}

Muons produced in the Earth could be a potential source of background
for the NLSPs events. They can however be eliminated by requiring two
charged tracks in the detector. The remaining important background is
di-muon events originated from charm decay. Charm is produced
from high energy neutrino interactions through the following process: 
$\nu N \rightarrow \mu^- H_c \rightarrow \mu^- \mu^+ H_x \nu$,
where the charm hadron decays according to $H_c \rightarrow H_x \mu^+\nu$
and $H_x$ can be a strange or non-strange quark.

We computed the charm production cross section 
$\sigma(\nu N \rightarrow cX)$ from a d or s quark~\cite{abc2}.
Although this cross section is around an order of magnitude greater than
the one for NLSP production~\cite{abc2}, muons lose much more energy while 
traveling through the Earth
and therefore have to be produced very close to the detector. For this
reason their track separation in the detector will also be much smaller.

We performed a Monte Carlo simulation of the di-muon production, propagation 
and detection using the same procedure as for the NLSP.

\section{Results}

\begin{figure}[t]
\centering
\epsfxsize=200pt \epsfbox{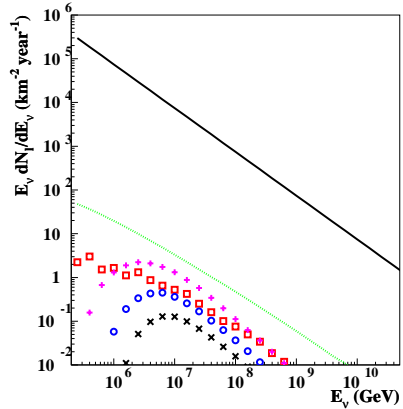}
\\*[-1.5cm]
\caption{Energy distribution of $\tilde\ell_R$ pair events per 
km$^2$, per year, at the detector. 
Curves that do not reach y axis; from top to bottom: 
$m_{\tilde q}=300$, ~$600$ and $900$~GeV. 
Here, $m_{\tilde\ell_R}=150$~GeV and
$m_{\tilde w}=250$~GeV. 
Also shown are the neutrino flux at earth and the $\mu$ and
the di-muon flux through the detector (curves that reach y axis; from top to bottom
respectively). In all cases we make use of the WB limit 
for the neutrino flux. 
}
\label{fig:rate}
%
\end{figure}

The energy distribution of NLSPs and di-muon background are shown in
Figure~\ref{fig:rate}. The number of events per km$^2$ per year is 
shown in Table~\ref{tab:nev} where not only a neutrino
incoming rate equal to the WB limit is assumed but also one
equal to the Mannheim, Protheroe and Rachen (MPR) limit~\cite{mpr}.

Figure~\ref{fig:rate} and Table~\ref{tab:nev} show that km$^3$
neutrino telescopes are sensitive to NLSP detection. Although the
number of di-muon events is larger than the NLSP rate there are
many ways to reduce this background.

\begin{table}
\begin{center}
\begin{tabular}{l|cccc}
\hline
\hline
 & $\mu^+\mu^-$ & $m_{\tilde{q}}=300$ & $600$ & $900$ (GeV) \\
\hline
WB & 30 & 6 & 1 & 0.3 \\
MPR & 1412 & 21 & 3 & 1 \\
\hline
\hline
\end{tabular}
\vspace*{.5cm}
\caption{\scriptsize Number of events per km$^2$ per year assuming the WB and MPR limits. The first column refers to 
di-muon events. The last three columns correspond to NLSP pair events, 
for three different choices of squark masses: $300$~GeV, $600$~GeV and $900$~GeV.
The number of di-muon events are given for energies
above $10^3$ GeV and of NLSPs above threshold for production.}
\label{tab:nev}
\end{center}
\end{table}

\begin{figure}[t]
\begin{center}
\epsfig{file=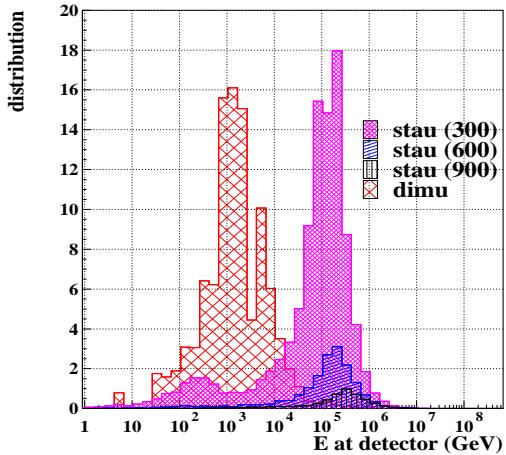,width=9.cm,height=8.cm,angle=0}
\\*[-1.cm]
\caption{Arrival energy distribution of the $\tilde\ell_R$ at the 
detector and for $m_{\tilde q}=300,~600 {\rm ~and~} 900$~GeV..
Here, $m_{\tilde\ell_R}=150$~GeV and
$m_{\tilde w}=250$~GeV. 
Also shown is the arrival distribution for the di-muon
background. The energy {\em deposited} in the detector by a stau 
traveling the average track length of $800~$m~\cite{als} is
$E_{\tilde{\ell}_R}^{\rm dep}= 150~$GeV, approximately the same for
all the masses considered here.
}
\label{fig:endist}
\end{center}
\end{figure}

First of all the lower integration limit to determine the number of events shown 
in Table~\ref{tab:nev} is different for the NLSP and the background. It is the
NLSP production threshold energy for the number of NLSPs and $10^3$ GeV
for the di-muon background.
The reason for this is that as the NLSPs lose less energy than
muons, they will look like lower energy muons in the detector. The NLSP
energy deposition in the detector is 150 GeV (see figure~( in \cite{abc2}).
For the same reason, one can cut higher energy events removing events
that deposit more than 300 GeV and reduce most of the dimuon
background. Figure~\ref{fig:endist} shows the arrival energy at the
detector for both NLSP and di-muon background where the efficiency of such
a cut is clear.

\begin{figure}[t]
%
\begin{center}
\epsfig{file=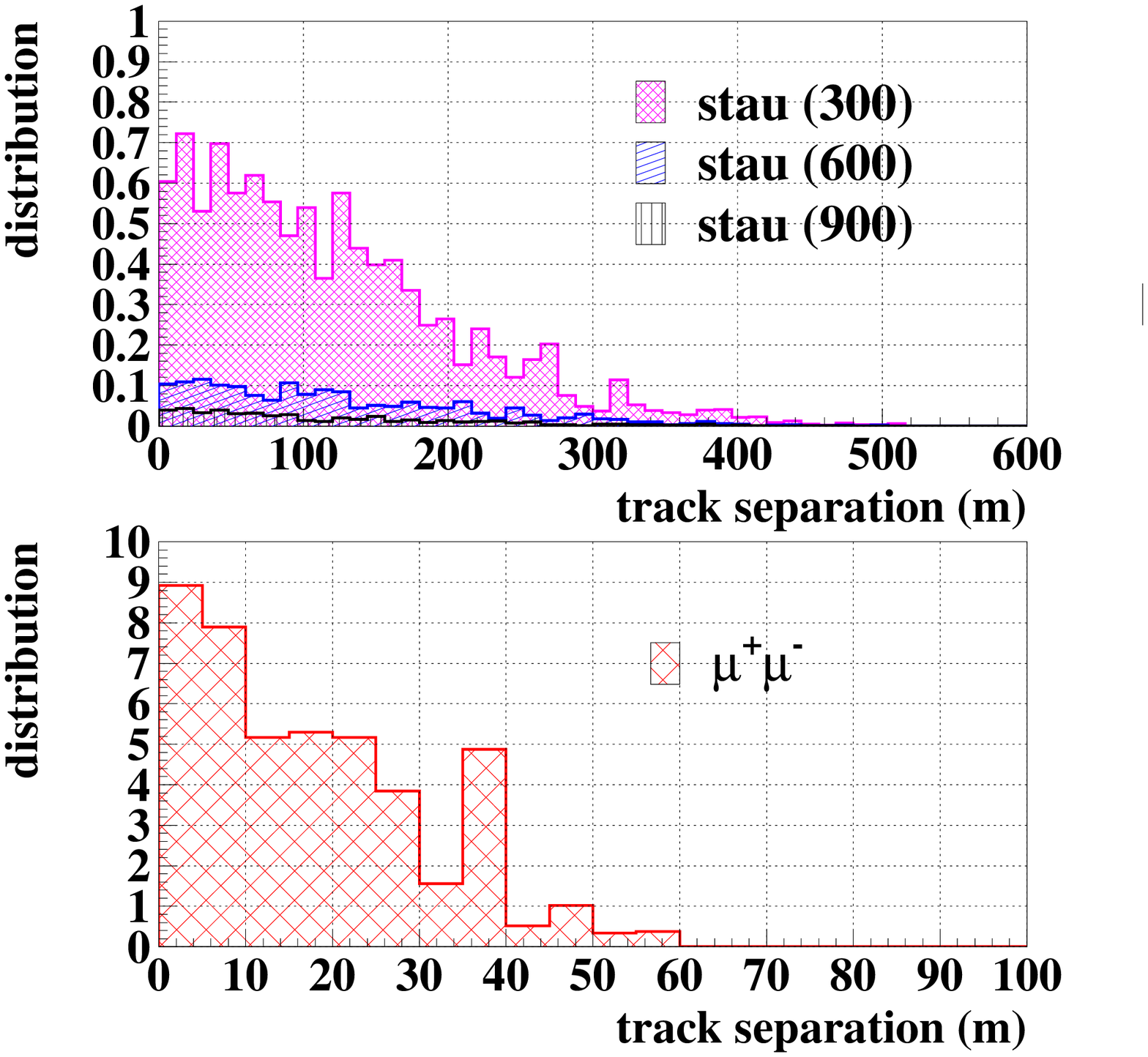,width=7.2cm,height=9cm,angle=0}
\\*[-1.cm]
\caption{Top panel: Track separation distribution of $\tilde\ell_R$
pair events. 
Bottom panel: The track separation distribution of the di-muon
background.  The relative 
normalization corresponds to the relative number of events for signal 
an background. Note the different horizontal scales, as well as
different binning between the two figures.
}
\label{fig:sep}
\end{center}
\end{figure}

Another powerful way for background reduction is the track separation between
2 NLSPs in the detector. This is shown in Figure~\ref{fig:sep}. While
a good fraction of the NLSP events are more than 50 m separated the dimuons
are less than 50 m separated. 

\section{Conclusions}

We conclude that Km$^3$ neutrino telescopes have the potential to discover
the NLSP if susy models that predict the gravitino as the LSP and
a charged slepton as the NLSP are correct. This will be an indirect 
determination of the dark matter. If the NLSP is observed it will constitute
a direct probe of the susy breaking scale. This search is complementary to
the LHC.

\section*{Acknowledgements}
{\small I.F.M.A participation in the EPNT workshop was partially supported
by the Brazilian National Counsel for Technological and Scientific 
Development (CNPq).}

\setcounter{section}{0}
\setcounter{subsection}{0}
\setcounter{figure}{0}
\setcounter{table}{0}
\newpage
\end{document}